# Comment on "Comment on 'Observation of a Push Force on the End Face of a Nanometer Silica Filament Exerted by Outgoing Light' "

Huakang Yu, Wei Fang, and Limin Tong*

State Key Laboratory of Modern Optical Instrumentation, Department of Optical Engineering,

Zhejiang University, Hangzhou 310027, China

* phytong@zju.edu.cn

In a recent comment [1], Masud Mansuripur presented some concerns with theoretical and experimental interpretations of She's paper [2], and concluded that She's conclusion is incorrect. We agree that, some of the She's work required further interpretation in detail. However, after careful investigation, we found that points raised in Mansuripur's comment are incomplete or incorrect, as explained below.

(1) About the emergent momentum in longitudinal direction: Mansuripur estimated the longitudinal momentum using a point-dipole oscillator model [1], and conclude that the light exiting the nanofiber carries only 75% of the momentum along the longitudinal direction. However, this is incorrect. Since a waveguiding nanofiber supports steady guiding modes that leaving a high fraction of evanescent waves outside the fiber core [3], the divergence of the output pattern [4] is much lower than that of a point-dipole oscillator. Using a three-dimensional FDTD simulation for a 450-nm-diameter silica nanofiber with refractive index 1.46 at wavelength 650 nm and a 520-nm-diameter silica nanofiber with refractive index 1.45 at wavelength 980 nm, we found that, in both cases the light exiting the nanofiber carries higher than 90% of the momentum along the longitudinal direction (measured 5.5 μm away from the output endface). Therefore, although the fiber diameter is smaller than the wavelength, the deviation of momentum of the output light due to the diffraction is relatively small.

(2) We agree that the group index $n_g$ should be used in Abraham momentum. The chromatic dispersion of nanofiber strongly depends on the fiber diameter [3]. We would like to point out that by properly choosing fiber parameters (e.g., diameter) and wavelength of the light, it is possible to have the group velocity equals to the phase velocity in a subwavelength-diameter fiber[3]. For reference, when operated at wavelength 980 nm, a 515-nm-diameter silica nanofiber offers $v_p = v_g = 0.685c$, and in She's work [2], the using of 520-nm-diameter silica nanofiber at 980-nm wavelength provided an approximately equal value of $v_p$ and $v_g$.

(3) About the mechanical momentum $p_{mech}$: after careful calculations, we conclude that $p_{mech}$ is only nontrivial for ultrafast pulses; for CW light used in She's work [2], the time averaging $p_{mech}$ along longitudinal direction is zero, as shown below. Thus the results in She's paper are still valid.

The mechanical momentum $p_{mech}$ can be obtained by Lorentz force, with the instantaneous Lorentz force density given by [5],

$$\mathbf{f} = (\mathbf{P} \cdot \nabla)\mathbf{E} + \frac{\partial \mathbf{P}}{\partial t} \times \mu_0 \mathbf{H}, \qquad (1)$$

where **E** and **H** are respectively the electric and magnetic fields of the optical mode in the nanofiber, **P** is the electric polarization density, and $\mu_0$ is the permeability of vacuum. In She's experiment, only the longitudinal Lorentz force density $\mathbf{f}_z$ is interested, i.e.,

$$\mathbf{f}_z = (\mathbf{P}\cdot\nabla)\mathbf{E}_z + (\frac{\partial \mathbf{P}}{\partial t}\times \mu_0 \mathbf{H})_z. \qquad (2)$$

Thus an arbitrary volume element Δv of the nanofiber will experience a longitudinal force $\mathbf{f}_z\Delta v$. For CW light, the mechanical momentum along longitudinal direction gained by Δv can be represented by integral over one optical period $T$,

$$p^z_{mech} = \Delta v \int_0^T \mathbf{f}_z dt. \qquad (3)$$

Take the fundamental optical mode ($HE_{11}$) for instance, we have electric and magnetic field components [6]:

$$\mathbf{E}_t = \hat{\mathbf{t}} E_t \cos(\beta z - \omega t) = (\hat{\mathbf{r}} E_r + \hat{\boldsymbol{\varphi}} E_\varphi)\cos(\beta z - \omega t), \qquad (4)$$

$$\mathbf{H}_t = \hat{\mathbf{t}} H_t \cos(\beta z - \omega t) = (\hat{\mathbf{r}} H_r + \hat{\boldsymbol{\varphi}} H_\varphi)\cos(\beta z - \omega t), \qquad (5)$$

$$\mathbf{E}_z = \hat{\mathbf{z}} E_z \sin(\beta z - \omega t), \qquad (6)$$

$$\mathbf{H}_z = \hat{\mathbf{z}} H_z \sin(\beta z - \omega t), \qquad (7)$$

$$\mathbf{P} = \varepsilon_0 (\varepsilon_r - 1)\mathbf{E}, \qquad (8)$$

Where $\hat{\mathbf{t}}$, $\hat{\mathbf{z}}$, $\hat{\mathbf{r}}$ and $\hat{\boldsymbol{\varphi}}$ are unit vector along transversal, longitudinal, radial and azimuthal direction, respectively; $E_t$ ($H_t$), $E_z$ ($H_z$), $E_r$ ($H_r$) and $E_\varphi$ ($H_\varphi$) are the transversal, longitudinal, radial and azimuthal electric (magnetic) fields, respectively; $\beta$ is the propagation constant; $\omega$ is the angular frequency of the field; $\varepsilon_0$ and $\varepsilon_r$ are the permittivity of vacuum and relative permittivity, respectively. With Eq. (4)-(8) and Eq. (2), the longitudinal Lorentz force density becomes,

$$\mathbf{f}_z = \hat{\mathbf{z}}\frac{1}{2}\varepsilon_0(\varepsilon_r - 1)\left[\left(E_r \frac{\partial}{\partial r}E_z + E_\varphi \frac{1}{r}\frac{\partial}{\partial \varphi}E_z + \beta E_z^2\right) + \mu_0 \omega \left(E_r H_\varphi + H_r E_\varphi\right)\right]\sin 2(\beta z - \omega t). \qquad (9)$$

By substituting Eq. (9) into Eq. (3), the time averaging longitudinal mechanical momentum $p^z_{mech}$ inside the nanofiber vanishes, i.e., $p^z_{mech}=0$ for CW light. Similar results can be obtained for higher order modes propagating in the nanofiber. This means the mechanical momentum would not exert net push force on the nanofiber endface.

References
[1] M. Mansuripur, Phys. Rev. Lett. **103**, 019301 (2009).
[2] W. She, J. Yu, and R. Feng, Phys. Rev. Lett. **101**, 243601 (2008).
[3] L. M. Tong, J. Y. Lou, and E. Mazur, Opt. Express **12**, 1026 (2004).
[4] S. S. Wang, J. Fu, M. Qiu, K. J. Huang, Z. Ma, and L. M. Tong, Opt. Express 16, 8887 (2008).
[5] R. Loudon and S. M. Barnett, Opt. Express **14**, 11855 (2006).
[6] A. W. Snyder, and J. D. Love, *Optical waveguide theory*, Chapman and Hall, New York, 1983.